\makeatletter

\newcommand{\Rmnum}[1]{\expandafter\@slowromancap\romannumeral #1@}
\makeatother
\documentclass[twocolumn]{aastex631}
\usepackage{amsmath}
\usepackage{graphicx}
\usepackage{amssymb}
\usepackage[utf8]{inputenc}

\begin{document}
	
\title{The detection prospects of the polarizations in the plateau phase of GRB afterglow by eXTP}
	
\author{Shu-Ting Wu}
\author[0000-0001-5641-2598]{Mi-Xiang Lan}
\author{Zelin Ren}
\affiliation{Center for Theoretical Physics and College of Physics, Jilin University, Changchun, 130012, China; lanmixiang@jlu.edu.cn}
\author[0000-0002-3776-4536]{Ming-Yu Ge}
\affiliation{Key Laboratory of Particle Astrophysics, Institute of High Energy Physics, Chinese Academy of Sciences, Beijing 100049, China}
\author[0000-0002-6299-1263]{Xue-Feng Wu}
\affiliation{Purple Mountain Observatory, Chinese Academy of Sciences, Nanjing 210008, China; xfwu@pmo.ac.cn}
\author[0000-0002-7835-8585]{Zi-Gao Dai}
\affiliation{Department of Astronomy, School of Physical Sciences, University of Science and Technology of China, Hefei 230026, China; daizg@ustc.edu.cn}

\begin{abstract}
Approximately (20-50)$\%$ of the gamma-ray burst (GRB) X-ray afterglows exhibit the shallow decay features. Two popular energy-injection models had been proposed to interpret such observational phenomenons, the relativistic wind bubble (RWB) model with a Poynting-flux injection and the structured ejecta (SE) model with a dynamical energy injection. Polarization predictions of the two models had been investigated and can be used as a test of the two models. However, the impacts of the parameters on the model predictions were not studied and the comparisons with the detection ability of the forthcoming mission, enhanced X-ray Timing and Polarimetry (eXTP), had not been discussed. We considered the above issues and found that influences of the model parameters on the predicted polarizations of the two models are very limited. To perform a feasible polarization detection during the plateau phase, the priority ToO response is required. The detection probability of the GRB plateau phase is about $1/3$ for one pointing under the priority ToO. The polarization detection probability would depend on the ratio between the Poynting-flux injection to the dynamical energy injection, which is unclear currently. The predicted flux density and polarization degree (PD) of the RWB model could be well above the threshold flux and minimal detectable polarization degree of the polarimetry focusing array (PFA) on board eXTP, while the predicted PDs of the SE model would be difficult to be detected by eXTP/PFA. Therefore, a detection of a significant polarization signal during the GRB plateau phase would prefer the RWB model and the injected energy would be in the form of the Poynting flux, while a non detection of the polarized signal would indicate a dynamical energy injection of the SE model.
\end{abstract}
\keywords{gamma-ray burst, polarization, eXTP}

\section{Introduction}
Gamma-ray bursts (GRBs) are among the most energetic phenomena in the universe \citep{1968JGR....73..434C,1973ApJ...182L..85K,1974ZhPmR..19..126M}, which results from the collapse of a massive star or the merger of two compact stars \citep{1992ApJ...395L..83N,1992AIPC..265...61M, 1993ApJ...405..273W, 2001ARep...45..236L, 2002ApJ...571..876Z,2003ApJ...591..288H,2017ApJ...848L..13A}. They are characterized by an initial prompt emission followed by a multi-stage afterglow observed at multiple wavelengths \citep{2006ApJ...642..354Z,2018pgrb.book.....Z}. The GRB afterglow is widely believed to originate from an external shock. The physical picture is that the outward-propagating shell interacts with the interstellar medium (ISM), generating a forward shock propagating into the medium and a reverse shock propagating into the shell. The synchrotron photons radiated by the shock-accelerated electrons would explain the typical power-law decaying light curves of the GRBs \citep{1997ApJ...485L...5W,1998PhRvL..81.4301D,2015ApJ...805...88Y}.

With the accumulation of the observational data, especially from the \emph{Swift} satellite, a significant proportion ($20\%-50\%$) \citep{2024A&A...692A..73G} of the GRB afterglows exhibit a X-ray plateau within $10^2$–$10^4$ s, during which the light curve flattens or decays shallowly, before transiting to the normal decay \citep{2014arXiv1401.1601Y,2019ApJ...883...97Z,2019ApJS..245....1T,2022ApJ...924...69Y}.

Continuous energy injection may exist to maintain the shallowly decay. There are two popular energy injection models of the GRB shallow decay phase. One is the relativistic wind bubble (RWB) model with the injected energy in the form of the Poynting flux \citep{2004ApJ...606.1000D,2007A&A...470..119Y}. \cite{2007ApJ...671..637Y} gives a detailed calculation of the high-energy emission, including the IC/SSC effects, in both the forward and shocked-wind regions under energy injection. Another energy injection model is the structured ejecta (SE) model with the injected energy in the form of dynamical energy \citep {1998ApJ...496L...1R, 2000ApJ...535L..33S}. The polarizations of the two models are predicted and it was found that the two models are testable with the X-ray polarization detections \citep{2016ApJ...826..128L}. However, there is non X-ray polarization detection during GRB plateau phase so far.

The enhanced X-ray Timing and Polarimetry (eXTP) mission \citep{2025arXiv250608101Z}, scheduled to be launched in early 2030, is a flagship mission dedicate to explore the X-ray astronomy through the time-domain, frequency-domain and polarizations \citep{2025arXiv250608101Z,2025arXiv250608369G,2025arXiv250608105B,2025arXiv250608104L,2025arXiv250608368Y,2025arXiv250608367Z}. The polarimetry focusing array (PFA) instrument on board eXTP satellite would focus on the polarization measurements \citep{2025arXiv250608101Z,2022ApJ...934..109Q,2023ExA....56..517Q}. The estimated event rate of the GRB plateau phase is about one per day \citep{2024A&A...692A..73G,2025arXiv250608368Y}. Depending on the models, the predicted PD could be as high as $60\%$ during the GRB X-ray plateau phase \citep{2016ApJ...826..128L}, which would be a promising target of the polarization detection.

In this paper, we revisit the RWB and the SE models to explore the influence of the model parameters on the polarization predictions. Then, based on the observational performance of the eXTP, its detection prospects are predicted. This paper is arranged as follows: In Section 2, we briefly introduce the two models. In Section 3, the impacts of the model parameters on the predicted polarizations are studied. In Section 4, the detection prospects by eXTP are discussed. Finally, in Section 5, conclusions and discussion are presented.

\section{The models}
The RWB model and the SE model are two widely discussed energy injection models in the shallow decay phase of the GRB afterglow \citep{2004ApJ...606.1000D,2000ApJ...535L..33S,1998ApJ...496L...1R}. In the following, we will briefly introduce the two models.

\subsection{The RWB model}
The RWB model was first proposed in \cite{2004ApJ...606.1000D}, which hypothesises that highly magnetised, rapidly rotating neutron stars are the central energy sources of the GRBs. Magnetar may undergo a direct loss of the rotational energy through rotationally driven processes. A relativistic wind which is predominantly characterised by an $e^+e^-$ pairs energy flux is ejected. This wind interacts with the outward-expanding fireball, forming a RWB. The wind bubble is comprised of two shocks: a reverse shock propagating into the cold wind and a forward shock propagating into the surrounding medium. These two shocks divide the space into four regions: (1) the unshocked interstellar medium (ISM), (2) the forward-shocked ISM, (3) the reverse-shocked wind gas, and (4) the unshocked cold wind.

There are two critical times of the RWB model. When the self-similarity parameter $\chi$\ equals to 1, the corresponding time is defined as $t_{cr}$ \citep{1976PhFl...19.1130B}. At this time, the injected energy can be compared with the initial energy $E_{0}$ of the outer shell.
\begin{equation}
    \label{equation1}
    t_{cr} = 4.9E_{52}L^{-1}_{w,47}\quad days,
\end{equation}
where $E_{52} = E_{0}/10^{52}$ \,ergs and the luminosity of the relativistic wind $L_{w,47}=L_{w}/10^{47} \,ergs \, s^{-1}$. Another critical time of the RWB model is $T_{M,0}$, which is the initial spin-down timescale of the magnetar.
\begin{equation}
    \label{equation2}
    T_{M,0} = 0.58B_{\perp,14}^{-2}I_{45}R^{-6}_{M,6}P^2_{0,ms} \quad days,
\end{equation}
where $I_{45}=I/10^{45}$ is the moment of inertia, $P_0$ is the rotation period, and $R_{M,6}= R_M/10^{6}$ is the radius of a millisecond magnetar.

The evolution of the RWB can be divided into three distinct phases: \Rmnum{1}. the initial-energy-dominated phase (\(t < t_{cr}\) ), the injected energy is less than the initial energy $E_0$ of the shell, and the energy of the system is dominated by $E_0$. The Lorentz factor of the forward shock region would evolve as \(\gamma_2 \propto t^{-3/8} \) and the Lorentz factor of the reverse shock is \(\gamma_3 \propto t^{-39/136} \). At this stage, the self-similarity parameter $\chi$ would be larger than 1. \Rmnum{2}. the injected-energy-dominated phase (\(t_{cr}< t < T_{M,0}\) ), the system's energy would be dominated by the injection. Both Lorentz factors $\gamma_2$ and $\gamma_3$ would evolve as $t^{-1/4}$. When \(t > t_{cr}\), $\chi = 1$. and \Rmnum{3}. the self-similar-evolution phase (\(t > T_{M,0}\)), the rotational speed of the magnetar has been slowed significantly, and the injected energy would be negligible. Hence the Lorentz factor of the forward shock would again be \(\gamma_2\propto t^{-3/8} \) and \(\ gamma_3\propto t^{-7/16} \) \citep{2004ApJ...606.1000D,1976PhFl...19.1130B,2000ApJ...545..807K}.

The polarisation evolution of the RWB model was studied in \cite{2016ApJ...826..128L}. The polarisation degree (PD) of the system is calculated according to \(\Pi = [(Q_{\nu,2}+Q_{\nu,3})^2 + U_{\nu,3}^2]^{1/2}/(F_{\nu,2}+F_{\nu,3})\), and the polarisation angle (PA) is formulated as \(\chi = \frac{1}{2}arctan( U_{\nu,3}/ (Q_{\nu,2}+ Q_{\nu,3})\), where $Q_{\nu,i}$, $U_{\nu,i}$, $F_{\nu,i}$ are the Stokes parameters of Region $i$.

\subsection{The SE model}
The SE model was first proposed by \citep{2000ApJ...535L..33S,1998ApJ...496L...1R}. The initial ejecta, which powers the prompt GRB, propagate into a wind environment with a density profile $n\propto R^{-g}$. The injected mass, whose velocity is faster than that with a Lorentz factor of $\gamma$, is $M(>\gamma)\propto\gamma^{-s}$ with $s>1$. After the deceleration of the initial ejecta, the injected mass will gradually collide with the outer ejecta and energise it. There are also two critical times of the system: the deceleration time $t_{0}$ of the initial ejecta and the injection termination time $t_{end}$.
\begin{equation}
    \label{equation3}
    t_{0} = 0.5R_0c^{-1}\gamma_0^{-2}
\end{equation}
where $R_0$ is the deceleration radius and initial Lorentz factor is denoted as $\gamma_0$.

There are still three stages. When \(t < t_0\), the injection is unimportant and the dynamics of the system is described as follows.
\begin{equation}
    \label{equation4}
     \gamma = \gamma_0,\quad R = 2c\gamma^2t
\end{equation}
When \(t_0 < t < t_{end}\), the bulk Lorentz factor will evolve as,
\begin{equation}
    \label{equation5}
     \gamma_2 = \gamma_3 = \gamma_0(t/t_0)^{-(3-g)/(7+s-2g)}
\end{equation}
\begin{equation}
    \label{equation6}
     R =R_0(t/t_0)^{(1+s)/(7+s-2g)}
\end{equation}
When \(t>t_{end}\), energy injection ends and the evolution of the afterglow will follow the standard non-refreshed model.
\begin{equation}
    \label{equation7}
     \gamma_2 \propto t^{-3/8},\quad \gamma_3 \propto t^{-7/16},\quad R \propto t^{-1/4}
\end{equation}

Since the central engine for the SE model would be a black hole, there are two possible magnetic field configurations in the injected material. One is the toroidal configuration, which might be carried out from the central engine by the Blandford-Znejek mechanism \citep{1977MNRAS.179..433B}. The other is the random field, which may be generated or amplified by the shocks. Under these two configurations, the Stokes parameters U of the synchrotron radiation is zero. The PD of the emission can be expressed as \(\Pi=(Q_{\nu, fs}+Q_{\nu, rs})/(F_{\nu, fs}+F_{\nu, rs})\). The PA rotates abruptly by $90^\circ$ when the PD changes its sign.

\section{The impact of the model parameters}
In \cite{2016ApJ...826..128L}, the impacts of the model parameters on the predicted polarizations are not discussed, which would be important for estimating the detection possibility by eXTP. So in the following, we will investigate the influence of the parameters on the predicted polarizations of the two models (the RWB and SE models).

\subsection{The RWB model}
\begin{figure*}
\centering
\includegraphics[width=1.0\textwidth]{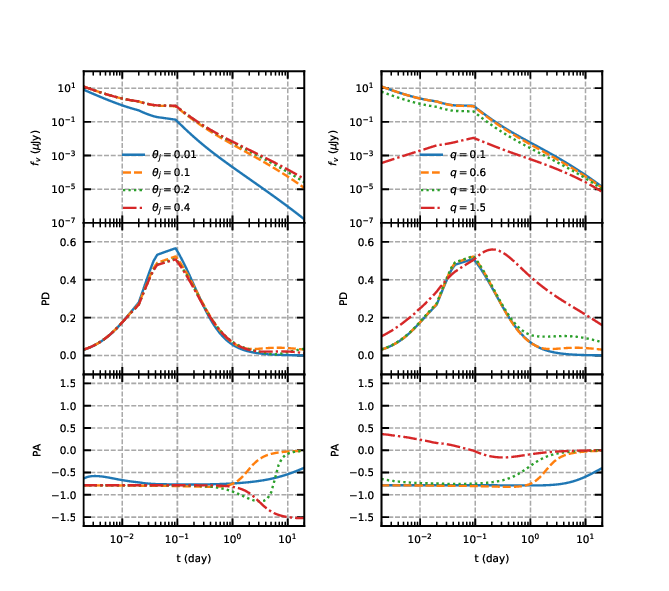}
\caption{The impact of the parameters on the light curves and polarisation evolutions of the RWB model at the eXTP/PFA 2$-$8 keV X-ray band. The influence of the jet half-opening angle (the viewing angle) is shown in the left (right) panel. The top, middle and bottom panels show the light curves, PD curves and PA curves, respectively. The blue-solid, orange-dashed, green-dotted and red-dash-dot lines in the left panel correspond to $\theta_j=$0.01 rad, 0.1 rad, 0.2 rad and 0.4 rad, respectively. The same sequence of the line styles in the right panel correspond to $q=$0.1, 0.6, 1.0 and 1.5.}
\label{fig:rwb}
\end{figure*}

For a magnetar central engine, most of its parameters would not vary significantly, so most of the model parameters can be set as fixed values. Unless otherwise stated, the following typical values of the parameters are taken: $E_{52} = 0.5$, the number density of the ISM $n_1 = 1cm^{-3}$, $\gamma_{w} = 10^4$, $L_{w,47} = 120$, $I_{45} = 2.4$, $P_0 = 1 ms$ and the indices of the energy spectra of the leptons $p_{3}= p_{2} = 2.5$. The energy sharing factors for electrons and the magnetic field in the forward shock region are $\varepsilon_{e,2} = \varepsilon_{b,2} = 0.1$, the energy sharing factor for $e^+ e^-$ in the reverse region is $\varepsilon_{e,3} = 0.9$ and the energy sharing factor for the magnetic field is $\varepsilon_{B,3} = 0.1$. The magnetic field in the injected wind is aligned at a direction of $\delta = \pi/4$. With our calculations, only two parameters (the jet half-opening angle $\theta_j$ and the observational angle $\theta_V$) show some limited influences on the polarizations. And the impacts of these two parameters on the light curves and polarizations are presented in \autoref{fig:rwb}.

In the left panel of \autoref{fig:rwb}, the observational angle is set as the typical value of $q = 2/3$ \citep{1999MNRAS.309L...7G} to study the influence of the jet half-opening angles on the light curves and polarisation curves. We find that for all the jet half-opening angles calculated, a shallow decay phase is clearly present. The flux density of a narrow jet \ ($\theta_j = 0.01$) rad is lower than that of the other calculated angles. The PD bump during the plateau phase, as expected, appears in all calculated jet half-opening angles. The PD of a narrow jet \ ($\theta_j = 0.01$) rad would be slightly higher than that for the wider ones. Then we set $\theta_j = 0.1$ rad to investigate the effects of the observational angle on the polarizations. For the off-axis observation with $q=1.5$, the plateau phase disappears and instead there is a peak in the light curve. And there is a PD bump slightly after the light curve peak.

\subsection{The SE model}

The light curves and polarization evolutions of the afterglow plateau phase under the SE model are also calculated and the impact of the parameters on the polarizations are studied. Unless otherwise stated, the parameters would take the following typical values: $E_{52} = 0.5$, the initial Lorentz factor $\gamma_0=125$, the jet half-opening angle $\theta_j=0.1$ rad, the viewing angle $q=2/3$, the electron spectral indices in the reverse shock region $p_{3}=2.5$ and in the forward shock region $p_{2}=2.5$. The energy equipartition factors for electrons and magnetic fields in the forward shock region are taken as $\varepsilon_{e, 2} = 0.02$ and $\varepsilon_{B, 2} = 0.001$, respectively. In the reverse shock region, the energy equipartition factor of the electrons is $\varepsilon_{e, 3} = 0.02$. For a random magnetic field, its energy equipartition factor is set as $\varepsilon_{B, 3} = 0.001$ and for an ordered toroidal magnetic field, it is $\varepsilon_{B, 3} = 0.1$. Here for simplicity we assume an ISM environment with particle number density $n_1 = 0.1$ and $g=0$. The end time of the energy injection is set at 0.3 days. The appearance of the plateau phase in X-ray bands of the SE model requires that the index $s$ should take a value of \(s = 3p_{2}-1\).

\begin{figure*}
\centering
\includegraphics[width=1\textwidth]{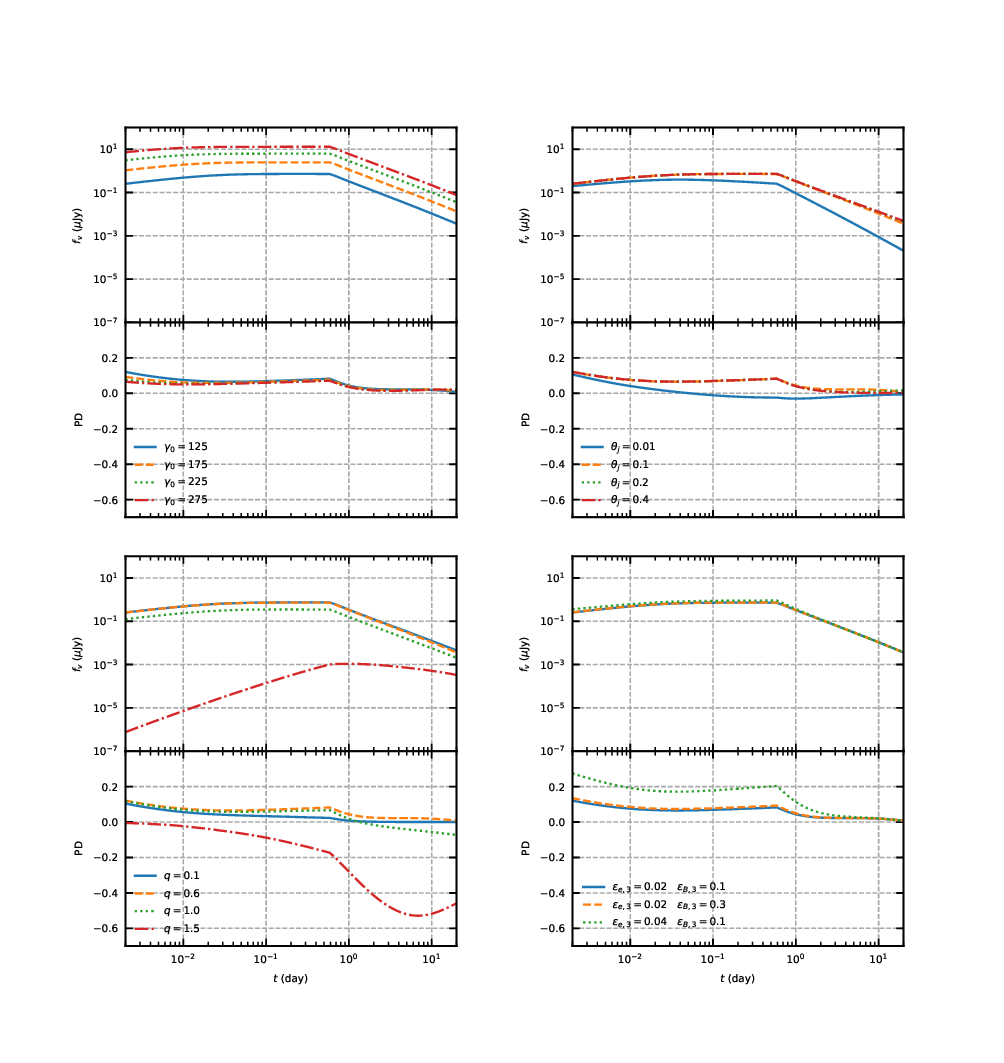}
\caption{The impact of the parameters on the light curves (upper panel) and PD curves (lower panel) of the SE model with a toroidal magnetic field in the reverse shock region at eXTP/PFA 2$-$8 keV X-ray band. The upper-left, upper-right, lower-left and lower-right panels show the impact of the initial Lorentz factor, the jet half-opening angle, the observational angle and the energy participation factors on the polarizations, respectively. In the upper-left panel, the blue-solid, orange-dashed, green-dotted and red-dash-dot lines correspond to $\gamma_0=$125, 175, 225 and 275, respectively. In the upper-right panel. the same sequence of line styles show $\theta_j=$0.01 rad, 0.1 rad, 0.2 rad and 0.4 rad. In the lower-left  panel, the same sequence of line styles is related to $q=$0.1, 0.6, 1.0 and 1.5. In the lower-right panel, the blue-solid, orange-dashed and green-dotted lines correspond to ($\varepsilon_{e,3}$, $\varepsilon_{B,3}$)=(0.02, 0.01), (0.02, 0.3) and (0.04, 0.01), respectively.}
\label{fig:tor}
\end{figure*}

The impact of the parameters on the light curves and polarisation evolutions of the SE model with a toroidal magnetic field in the reverse shock region at 2$-$8 keV of the eXTP/PFA band are presented in \autoref{fig:tor}. The values of the initial Lorentz factor do not affect the evolutionary trend of both the light curve and the PD curve. The PD of the SE model during the X-ray shallow decay phase maintain a constant value of approximately 7\%. However, the predicted flux density would increase significantly with initial Lorentz factor. The PD of the narrow jet ($\theta_j = 0.01$) is roughly 0 during the shallow decay phase, while it is 7\% for the wider jets. The impact of the jet half-opening angle on the light curve would be negligible. The plateau is present for the on-axis observation and disappears for the off-axis observation ($q=1.5$). The predicted PD during the plateau phase would be no larger than 7\% for various viewing angles. The influence of the energy participation factor of the electrons in the reverse shock region $\varepsilon_{e,3}$ on the predicted PD is large, because the contributions of the highly polarized emission from the reverse shock region would increase with $\varepsilon_{e,3}$. However, the impact of the $\varepsilon_{B,3}$ on the predicted PD is negligible. The predicted PD can reach $20\%$ with $\varepsilon_{e,3}=0.04$ and the flux density from the reverse shock region would be comparable with that from the forward shock region. Since the total flux density of the SE model would be dominated by the forward shock region, a $20\%$ PD would be the upper limit of the SE model.

\begin{figure*}
\centering
\includegraphics[width=1\textwidth]{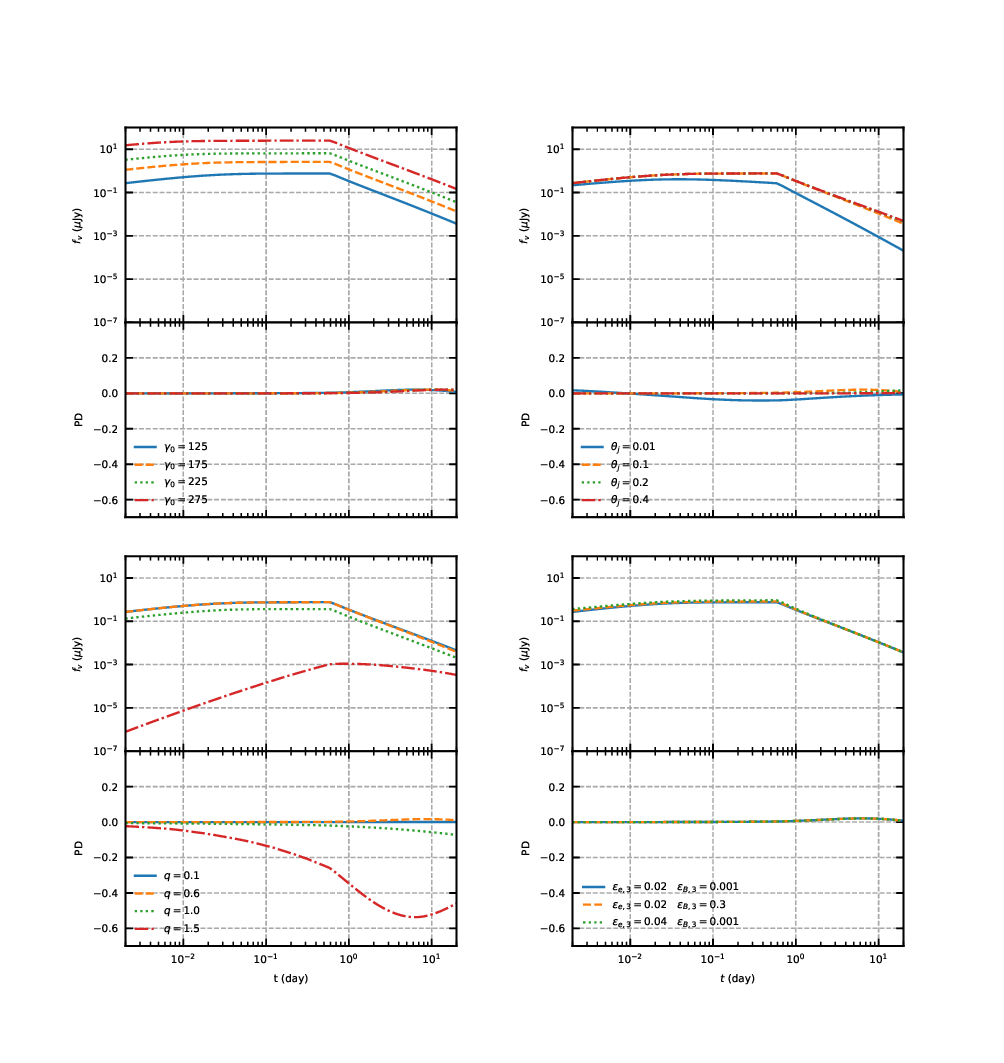}
\caption{Same as \autoref{fig:tor}, but with a random magnetic field in the reverse shock region.}
\label{fig:ra}
\end{figure*}

The impact of the parameters on the light curves and polarisation evolutions of the SE model with a random field in its reverse shock region at the eXTP/PFA 2$-$8 keV X-ray band are depicted in \autoref{fig:ra}. All five parameters ($\gamma_0$, $\theta_j$, $q$, $\varepsilon_{e,3}$ and $\varepsilon_{B,3}$) show negligible effect on the predicted $\sim0\%$ PD during the plateau phase.

\section{Detection prospects of eXTP}
Since the impact of the model parameters on the predicted polarizations of the two energy injection models of the plateau phase are roughly negligible, we can give model predictions with the typical values of the parameters (see Section 3) and compare them with the detection ability of the eXTP/PFA \citep{2025arXiv250608101Z}.

\begin{figure*}
\centering
\includegraphics[width=1.0\textwidth]{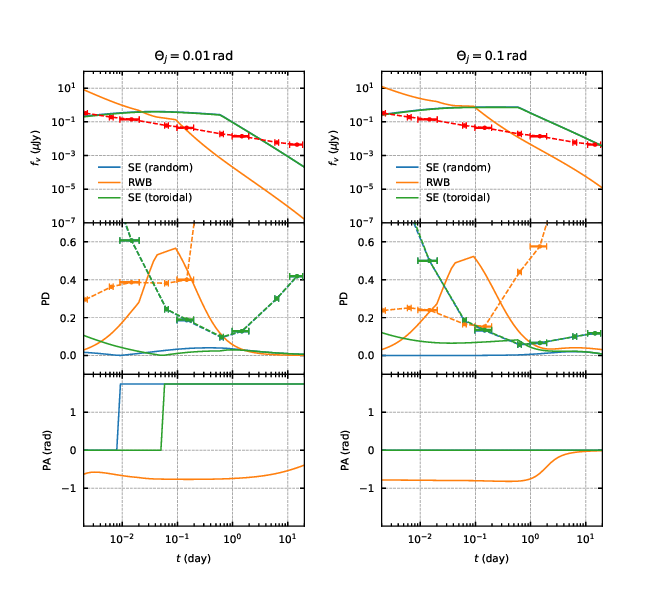}
\caption{The predicted Light curves and polarisation evolutions of the GRB plateau phase at 2$-$8 keV and their comparisons with the corresponding threshold flux and MDPs of eXTP/PFA. The top, middle and bottom panels show the light curves, PD curves and PA curves, respectively. The jet half-opening angle is 0.01 rad (0.1 rad) in the left (right) panel. The blue, green and orange solid lines represent the predictions of the SE model with a random field in the reverse shock region, SE model with a toroidal field in its reverse shock region and the RWB model. The red dashed lines in the top panel show the threshold flux of the eXTP/PFA. The blue-dashed, green-dash-dotted and orange-dashed lines represent the corresponding MDPs of the SE model with a random field, the SE model with a toroidal field and the RWB model.}
\label{fig:fv}
\end{figure*}

The light curves and polarisation evolutions at 2-8 keV energy band of the two models are compared with the threshold flux and the MDPs of the eXTP/PFA, which is shown in \autoref{fig:fv}. Two jet half-opening angles are considered with one narrow jet ($\theta_j=0.01$ rad) and another wider jet ($\theta_j=0.1$ rad). For the RWB model, both the predicted flux density and PD during the plateau phase are well above the corresponding threshold flux and MDP curves, respectively. However, even the predicted flux is well above the threshold of eXTP/PFA, the predicted PDs of the SE model would be difficult to be detected by the eXTP/PFA. The above model predictions are based on the results calculated with the typical parameters. If the proportion of the flux density from the reverse shock region with a toroidal field of the SE model becomes larger, the predicted PD of the jet emission would also increase. With our calculation, the predicted PD can reach $\sim20\%$ for the SE model with a toroidal field in its reverse shock region when the flux density from the reverse shock region are comparable with that from the forward shock region. Therefore, the detection of the PD during the GRB plateau phase would prefer the RWB model, while the SE model with a toroidal field in the reverse shock region could not be strictly rejected.

\begin{figure*}
\centering
\includegraphics[width=1.0\textwidth]{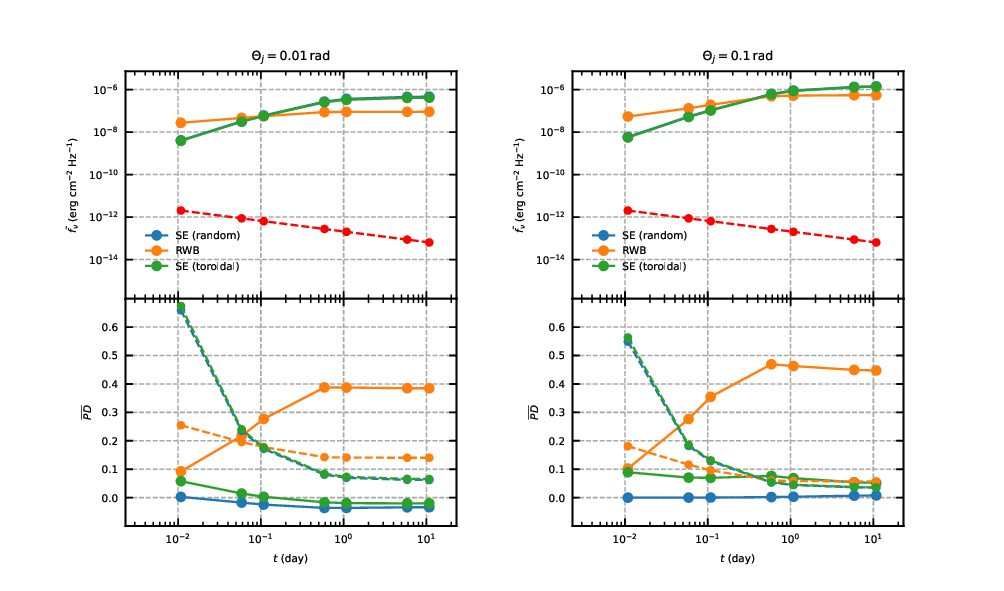}
\caption{Predicted time-accumulated light curves and polarization evolutions of GRB plateau phase at 2$-$8 keV and their comparisons with the corresponding threshold flux and MDPs of eXTP/PFA. The top and bottom panels show the light curves and PD curves, respectively. The left and right panels correspond to the jet half-opening angles of 0.01 rad and 0.1 rad, respectively. The blue, green, and orange solid curves represent the predictions of the SE model with a random magnetic field in the reverse shock region, the SE model with a toroidal magnetic field, and the RWB model, respectively. The red dashed line in the top panel indicates the threshold flux of eXTP/PFA. The blue, green, and orange dashed lines denote the MDPs for the SE model with a random field, the SE model with a toroidal-field, and RWB model, respectively. The circles represent our calculated points.}
\label{fig:inin}
\end{figure*}

\autoref{fig:inin} presents the evolution of the time-accumulated flux ($ \overline{f_v}$) and time-accumulated $ \overline{PD}$ as a function of the observation time $t$. The upper panel shows that $\overline{f_v}$ increases monotonically with time, and the resulting $ \overline{f_v}$ values for both RWB and SE models lie above the threshold flux of eXTP/PFA. In the lower panel, it is clearly shown that as the observation time increases, the $ \overline{PD}$ predicted by the RWB model becomes detectable by eXTP/PFA, whereas the $ \overline{PD}$ predicted by the SE model remains difficult to be detected.
\begin{equation}
    \overline{f_\nu} = \int_{0}^{t} f_{\nu}  \, dt_{\mathrm{obs}}
    \label{equation8}
\end{equation}
\begin{equation}
    \overline{Q_{\nu}} = \int_{0}^{t} Q_{\nu}  \, dt_{\mathrm{obs}}
    \label{equation9}
\end{equation}
\begin{equation}
     \overline{U_{\nu}} = \int_{0}^{t} U_{\nu}  \, dt_{\mathrm{obs}}
    \label{equation10}
\end{equation}
\begin{equation}
    \overline{PD} = \frac{\sqrt{\overline{Q_{\nu}}^2 + \overline{U_{\nu}}^2}}{\overline{f_{\nu}}}
    \label{equation11}
\end{equation}

Both the time-resolved and time-integrated spectra of the plateau phase under typical parameters are predicted at the 2-8 keV energy band and compared with the detection ability of the eXTP/PFA. Since the flux and the predicted PD during the plateau phase varies slightly with time, we simply select the time-resolved spectra at the half of the end time of the plateau phase ($t_b/2$). The time-integrated spectra are obtained via the integrations from the beginning time to the end time of the plateau phase at 2-8 keV X-ray band. Two jet half-opening angles are considered, one narrow jet with $\theta_j=0.01$ rad and another wider jet with $\theta_j=0.1$ rad. The predicted flux densities of the two models are well above the threshold of the eXTP/PFA, while only the predicted PD of the RWB model are high enough to be detected by eXTP/PFA.

\begin{figure*}
\centering
\includegraphics[width=1.0\textwidth]{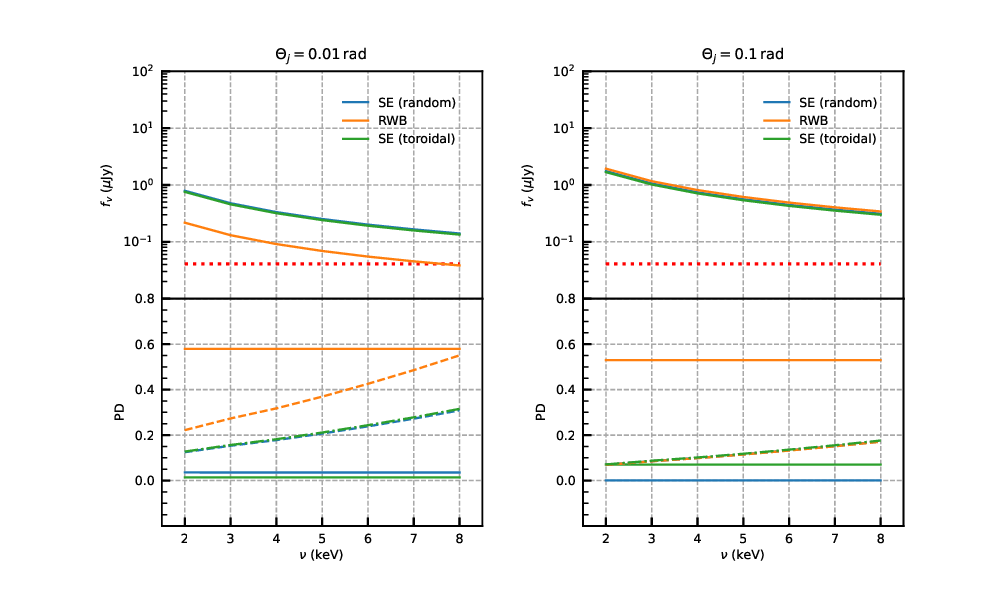}
\caption{The predicted time-resolved (at tb/2 and $t_b$ is the end time of the shallow decay phase) energy spectra (top panel) and PD spectra (bottom panel) at 2-8 keV compared with the corresponding threshold flux and MDPs of eXTP/PFA. The jet half-opening angle is 0.01 rad (0.1 rad) in the left (right) panel. The blue, green and orange solid lines show the predictions of the SE model with a random field in the reverse shock region, the SE model with a toroidal field in the reverse shock region and the RWB model, respectively. The red-dotted line in the top panel show the threshold flux of eXTP/PFA. The blue-dashed, green-dash-dotted and orange-dashed lines in the bottom panel show the corresponding MDPs of the SE model with a random field, the SE model with a toroidal field and the RWB model, respectively.}
\label{fig:tb}
\end{figure*}

\begin{figure*}
\centering
\includegraphics[width=1.0\textwidth]{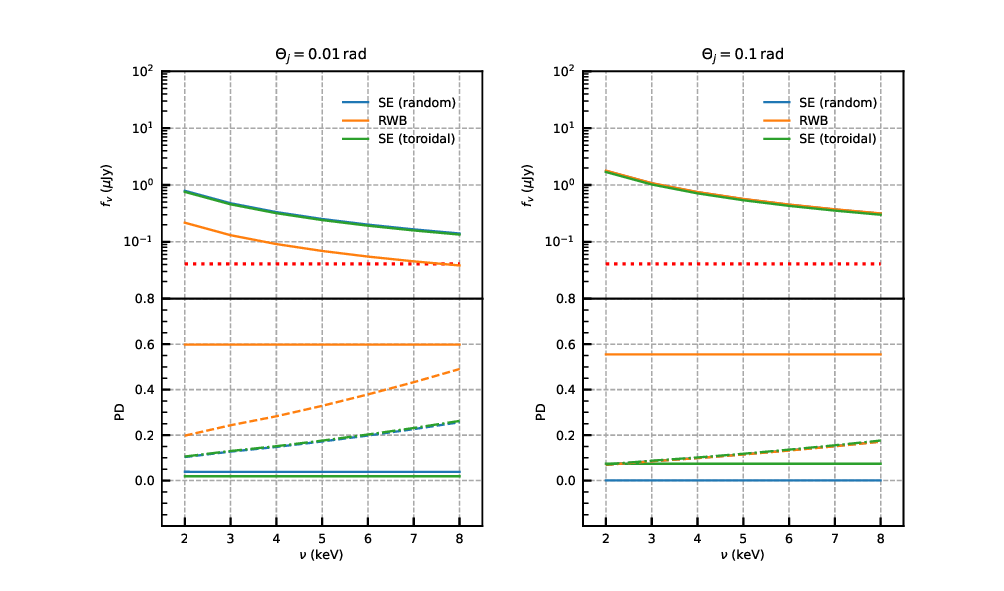}
\caption{Same as \autoref{fig:tb}, but for the time-integrated ones.}
\label{fig:in}
\end{figure*}

To clearly show the feasibility of X-ray polarization detection by eXTP during the plateau phase of GRB afterglow, we combine the expected Target of Opportunity (ToO) response capability of eXTP to provide a detailed estimation of the achievable polarization detection. The Wide-field Monitor (WFM) of eXTP can monitor one-quarter of the sky and can rapidly locate transient sources. Once a transient source (such as a GRB) is detected, the telescope can autonomously slew at a speed of 3$^\circ$ per minute and transmit the coordinates to the ground within 30 seconds via the BeiDou Satellite System. eXTP can initiate observations during the available contact window, typically achieving target pointing within 20 minutes (0.013 days) for priority ToOs and within 13 hours (0.542 days) for routine ToO \citep{2019SCPMA..6229506I,2025SCPMA..6819507Z,2025arXiv250608101Z,2025arXiv250608369G,2025arXiv250608105B,2025arXiv250608104L,2025arXiv250608368Y,2025arXiv250608367Z}. The detailed detection possibility of the GRB plateau phase can be found in \cite{2025arXiv250608368Y}. However, they did not consider the ToO response of eXTP.

Statistics indicate that approximately $35\%$ of the GRBs exhibit a plateau phase \citep{2024A&A...692A..73G}. Among the bursts with plateau phase, the fraction of the bursts with plateau end time ($t_b$) larger than 13 hours is only 16\%. And for the bursts with $t_b > 13$ hours, the fraction of the bursts with flux exceeding the PFA threshold is $94\%$ \citep{2019SCPMA..6229506I,2025SCPMA..6819507Z,2025arXiv250608101Z,2025arXiv250608369G,2025arXiv250608105B,2025arXiv250608104L,2025arXiv250608368Y,2025arXiv250608367Z}. As a result, the probability of successfully detecting the end of the plateau phase is only $5\%$ for one pointing under the routine ToO. More importantly, at the time of 13 hours, the RWB model predicts a maximum PD of 14\%, while the SE model predicts a PD close to $0\%$. Both values are below the corresponding MDP ($15.54\%$) of eXTP/PFA. Therefore, under these circumstances, eXTP cannot detect a polarization signal during the plateau phase with the routine ToO. In contrast, the priority ToO appears much more promising. Among the bursts with plateau phase, the fraction of the bursts with $t_b > 20$ minutes is $91\%$. And for the bursts with $t_b > 20$ minutes, the fraction of the bursts with flux exceeding the PFA threshold is 99\%. Therefore, the detection probability of the GRB plateau phase is $32\%$ for one pointing under the priority ToO. At 20 minutes after the burst trigger, the RWB model predicted a minimum PD of $23\%$, which is higher than the corresponding MDP ($22\%$) of PFA. However, the SE model predicts a maximum PD of only $7\%$, which remains below the MDP and thus cannot be detected by eXTP. Since the polarization detection probability would also depend on the ratio of the Poynting-flux injection to the dynamical energy injection, which is unknown currently, it is hard to make such an estimation.

\begin{table*}[ht]
\centering
\caption{Detection Capability of eXTP/PFA at Different post-trigger Times}
\label{tab:detection}
\begin{tabular}{cccccc}
\hline
Post-Trigger Time (day) &  $f_{{\nu}, typical} (\mu\text{Jy})$ & $MDP_{typical}$ & $PD_{RWB}$ & $PD_{SE, random}$ & $PD_{SE, toroidal}$ \\
\hline
0.004 & 9.750 & 0.191 & 0.078 & 0.009 & 0.072\\
0.05 & 0.706 & 0.200 & 0.496 & 0.024 & 0.001\\
0.1 & 0.315 & 0.212 & 0.497 & 0.034 & 0.012\\
\hline
\end{tabular}
\end{table*}

The plateau phase in the X-ray afterglow of GRBs typically begins at approximately $t_a$=($10^2 - 10^3$) s=(0.0012 - 0.012) days after the burst and ends at $t_b$=($10^{3.5} - 10^{4.5}$) s=(0.037 - 0.37) days \citep{2007ApJ...670..565L, 2014arXiv1401.1601Y,2019ApJ...883...97Z}. We selected three key observation times ($t_a = 0.004 \ day$, $t_b/2 = 0.05 \ day$, and $t_b = 0.1 \ day$) as shown in \autoref{tab:detection} to systematically evaluate the detection capability of the GRB plateau phase by eXTP/PFA. The observed typical flux at each time is significantly higher than the corresponding threshold flux of eXTP/PFA. In terms of polarization detection, the calculated MDP (based on the observed typical flux) are 0.191, 0.200 and 0.212. The PDs predicted by the SE model (0.009, 0.024, 0.034 for random magnetic field and 0.072, 0.001, 0.012 for toroidal magnetic field) are all lower than the corresponding MDP, failing to meet the detection requirements. In contrast, the PDs predicted by the RWB model at $t_b/2$ and $t_b$ are 0.496 and 0.523, which significantly exceed the corresponding MDP, indicating that eXTP/PFA can reliably detect and verify the polarization signals of RWB model. The response time of the priority ToO is shorter than the typical ending time of the plateau phase. The observed typical flux is well above the PFA threshold. And the predicted PDs of the RWB model are above the corresponding MDPs, while they are below the MDPs for the SE model. Therefore, the priority ToO could make a detection of the plateau phase with typical ending time. Since the observed typical ending time is shorter compared with the response time of the routine ToO, the GRB plateau phase with typical ending time would not be detectable for the routine ToO.

Furthermore, we use a complete cosmological physical framework to calculate the redshift detection threshold of eXTP/PFA. This framework is based on the standard $\Lambda$CDM cosmological model with parameters $H_0 = 69.3~\text{km}~\text{s}^{-1}~\text{Mpc}^{-1}$, $\Omega_m = 0.27$ and $\Omega_\Lambda = 0.73$ \citep{1992ARA&A..30..499C,hogg2000distancemeasurescosmology}, and it comprehensively incorporates key physical factors including the intrinsic luminosity of sources $L$, the minimum detectable flux $f_{min}$ and cosmological $K$-correction \citep{2001AJ....121.2879B,hogg2002kcorrection}. The luminosity $L$ is the rest-frame luminosity of the plateau phase \citep{2010ApJ...722L.215D}.
We adopted a bimodal luminosity distribution as input. The statistical maximum luminosity ($L_{\max} = 10^{48.38}~\text{erg}~\text{s}^{-1}$) represents the luminosity of the brightest GRB plateaus. The modal luminosity ($L_{\text{modal}} = 10^{47.64}~\text{erg}~\text{s}^{-1}$) reflects the typical luminosity of the GRB plateaus. The redshift threshold is determined by numerically solving the cosmological radiative transfer equation. To determine $f_{min}$, we adopted the larger one of two constraints ($f_{threshold}$ and $f_{pol}$). The $f_{threshold}$ is the threshold flux of eXTP/PFA, and $f_{pol}$ is the flux required to achieve a MDP, which equals to the predicted maximum PD ($PD_{max}$) of the model. For the RWB model, $PD_{max}$ is 0.57 at $t=0.08$ days. At the time of 0.08 days, we have $f_{threshold} = 6.36 \times 10^{-13} \text{ erg s}^{-1}$, while $f_{pol} = 7.89 \times 10^{-13} \text{ erg s}^{-1}$. So we set $f_{min} = 7.89 \times 10^{-13} \text{ erg s}^{-1}$ to ensure simultaneous detection of the flux and polarization. Similarly, for the SE model, $PD_{max}$ is 0.07 at $t=0.12$ days. The flux threshold is $f_{threshold} = 5.18 \times 10^{-13} \text{ erg s}^{-1}$, while the polarization requirement is $f_{pol} = 3.44 \times 10^{-11} \text{ erg s}^{-1}$. Therefore, we selected $f_{min} = 3.44 \times 10^{-11} \text{ erg s}^{-1}$.
For the brightest GRB plateaus, the detection limit of the redshift extends to $z_{max} \approx 12.62$ for the RWB model, whereas it is $z_{max} \approx 2.65$ for the SE model. For the GRB plateaus with typical luminosity, the limit is $z \approx 6.15$ for the RWB model and is $z \approx 1.35$ for the SE model, respectively.

\section{Conclusions and discussion}
In this paper, the impact of the parameters on the predicted polarizations of the GRB plateau phase under two popular models (the RWB and SE models) are discussed. Then the model predictions are compared with the detection ability of the eXTP/PAF \citep{2025arXiv250608101Z}.

Although approximately (20-50)\% of the \emph{Swift/XRT} observed X-ray afterglows exhibit the shallow decay phase \citep{2024A&A...692A..73G}, there is few polarization detection at X-ray band on this stage. Such detection would be important. Because the predicted PDs would be as high as $60\%$ for the RWB model during the plateau phase, while it is less than $20\%$ for the SE model \citep{2016ApJ...826..128L}. The polarization detection could distinguish these two popular models. However, the above conclusion are rather preliminary, since they did not consider the influence of the parameters on the predicted flux density and the polarizations. And also the model predictions are not compared with the detection ability of the corresponding detector.

Here, with the calculation we found two parameters (the jet half opening-angle $\theta_j$ and the observational angle $q$) of the RWB model and four parameters of the SE model (initial Lorentz factor $\gamma_0$, the jet half opening-angle $\theta_j$, the observational angle $q$ and the energy participation parameter of electrons in the reverse shock region $\varepsilon_{e,3}$) show some limited influence on the predicted PDs. When confronting with the detection ability of the eXTP/PFA, it is crucial to note that a feasible polarization detection during the GRB plateau phase requires the priority ToO response, as the predicted PDs of the two models both fall below the MDP of PFA after the response time of the routine ToO. The detection probability of the GRB plateau phase is roughly $1/3$ for one pointing under the priority ToO and the polarization detection probability would depend on the energy form of the injection. Since the fraction of the Poynting-flux injection during the plateau phase is unclear, it is hard to make such an estimation about the polarization detection probability. Under the priority ToO, the predicted flux density and PDs of the RWB model could be both well above the corresponding detection limit. While even the predicted flux density can be detected by eXTP/PFA, the predicted PDs of the SE model would be less than $20\%$ and with a typical value of $7\%$, which would be difficult to be detected by the eXTP/PFA. Therefore, under the priority ToO response, only the predicted PD of the RWB model could be detected by eXTP/PFA. If a significant PD is detected during the GRB plateau phase, the RWB model would be favored and the injected energy would be in the form of the Poynting flux. And a non-detection of the polarizations during the plateau phase would prefer the SE model and the injected energy would be in the form of the dynamical energy.

In addition, the polarization-detection-constrained redshift thresholds for both models are calculated under the $\Lambda$CDM cosmological framework. For the RWB model, the threshold is $z \approx 12.62$ for the most luminous plateau and is $z \approx 6.15$ for the plateaus with typical luminosity. For the SE model, the threshold is $z \approx 2.65$ for the most luminous plateau and is $z \approx 1.35$ for the plateaus with typical luminosity.

\section*{Acknowledgments}
We thank the anonymous referee for useful comments that improved our manuscript. This work is supported by the National Natural Science Foundation of China (grant No. 12473040, \ 12041306, \ 12393812).
M.X.L. would also like to acknowledge the financial support from Jilin University.

\twocolumngrid

\bibliography{ms_arxiv}

\end{document}